\documentstyle [epsf,epsfig]{article}
\begin{document}
\Large
\centerline {\bf  The reach of a future Linear Collider}
\centerline {\bf  after the g-2 result}
\vskip 1  cm
\centerline {F. RICHARD}
\smallskip
\centerline {\it Laboratoire de l'Acc\'el\'erateur Lin\'eaire, IN2P3-CNRS}
\centerline {\it et Universit\'e de Paris-Sud, F-91405 Orsay Cedex, France}
\vskip 1 cm
\normalsize
\begin{abstract}
 Combining the cosmological
requirement on dark matter 
with the recent BNL g-2 measurement  
it is argued that, within the mSUGRA framework,
the preferred region for SUSY mass parameters falls well inside the area 
covered by the future linear colliders under consideration
 for right handed sleptons and of the 2 lightest neutralinos. The coverage for the lightest chargino and left handed sleptons 
is also favoured but with smaller confidence.
\end{abstract}
 
\vskip 1 cm
The main uncertainty on the physics potential of future e$^+$e$^-$ colliders 
comes from our ignorance of the SUSY mass spectrum. As
recently pointed out in \cite{fengr}, if the mass spectrum of gauginos (charginos, neutralinos) and sleptons is light enough to be 
observable with a 
500 GeV LC, then signals should appear before LHC starts(2006)either in the precise measurement of g-2 under way at BNL,
or in the observation
of primordial neutralinos in the high sensitivity experiments under construction (CDMS, EDELWEISS, CRESST). \par
The recent indication \cite{bnl} 
reported from the BNL experiment although not yet conclusive (2.6 s.d. effect), is encouraging since it 
is compatible with the expected contributions of a light SUSY spectrum as will be discussed in more detail below. \par  
As already mentioned, there is no definite SUSY symmetry breaking, SSB, mechanism which can allow a precise prediction of the 
SUSY mass spectrum. In the most general approach this spectrum has more than 100 free parameters but 
there are various experimental constraints (in particular the requirement to avoid FCNC) which impose to reduce considerably
this number. \par
In the so-called gravity-mediated SSB, mSUGRA, there are 2 mass parameters, m$_0$ and m$_{1/2}$, one related to the mass of the
scalar superpartners and the other related to the mass of the fermionic superpartners,  and 3 mixing parameters tan$\beta$,
$\mu$ and A. These parameters enter in the definition of the SUSY particles, the Higgs bosons and the Z masses. 
The latter provides one constraint which allows to determine $\mu^2$, leaving unknown sign($\mu$). \par
Four ingredients play an important part in defining the allowed domain for m$_0$, 
m$_{1/2}$ and tan$\beta$ (A plays a less important role):
\begin{itemize}
\item The Higgs mass which, according to LEP, should be above 114 GeV (and probably at 115 GeV) which imposes m$_{1/2}$ 
above 350 GeV (this lower bound can be reduced to $\sim$ 200 GeV taking into account various uncertainties)
unless m$_0$ is very large (unlikely given the observation on g-2 as discussed below). 
\item The recent measurement of g-2 for the muons at BNL 
with a 2.6 sd excess above the SM value which, interpreted in terms of SUSY, crudely speaking says \cite{kane}
(at the 1.5 sd level) that $\tilde{m}$ is below 65$\sqrt{tan\beta}$ GeV, where $\tilde{m}$ is of order M$_{chargino}$+M$_{sneutrino}$.
From LEP2 we can tell that this quantity is above 200 GeV, implying that  tan$\beta$ should be larger than $\sim$ 10. 
The sign of g-2 also 
implies that $\mu$ should be positive.
\item The cosmological solution which favours small values of m$_0$ to generate the proper amount of dark matter in the 
universe($\Omega h^2\sim$0.3). 
To insure that the LSP is neutral (neutralino) and not charged (stau lepton) m$_0$ is limited from below. 
The co-annihilation process stau+neutralino, active when both particles have the same mass, favours 
solutions which are close to this limit.
\item The branching ratio b$\rightarrow$s$\gamma$ implies that for large tan$\beta$, m$_{1/2}$ and m$_0$ should be large to be 
compatible with the data. This constraint is only very restrictive for negative values of $\mu$ while for positive values, compatible
with the g-2 result, it simply implies that tan$\beta$ cannot be above $\sim$ 30 \cite{deboer}. 
\end{itemize}
Putting together the 4 requirements, one can easily derive a valid SUSY spectrum. Typical mass spectra for 3 relevant values of  
tan$\beta$ are given in table 1.
\begin{table}[t]
\centering
\caption{mSUGRA solutions }
\vskip 0.5 truecm
\begin{tabular}{|c|c|c|c|}
\hline
tan$\beta$ & 10 & 20 & 30 \\
\hline
m$_{1/2}$  & 400   &350   & 350   \\
\hline 
m$_0$ & 100   & 120 & 170  \\
\hline 
$\mu$ & 475 &415 & 415\\
\hline
$\delta a_{\mu}$ x 10$^{10}$  &14  & 33   & 43 \\
\hline
M $_{chargino1}$  &305  &265 & 265 \\
\hline
M$_{neutralino1}$ &160 &140 &140 \\
\hline
M$_{sleptonR}$ &190&186&221 \\
\hline
M$_{sneut}$  &395 &270 &295  \\
\hline
M$_{chargino2}$ &500  &440 &440 \\
\hline
M$_{neutralino2}$ &305 &265 &265 \\
\hline
M$_{squark}$ &900&800 & 800 \\
\hline
M$_{gluino}$ & 900 & 790 & 790 \\
\hline 
Rate/100kg/day Ge &0.05&0.25&0.7\\
\hline
\end{tabular}
\end{table} 
 Note that one can adjust the lightest Higgs mass to 115 GeV with the A parameter. 
\par
As often discussed in the litterature,             
the dark matter constraint suggesting light neutralinos and sleptons can be evaded in several
ways. Without discussing the details one may simply state that the g-2 result prevent most of these scenarios. One can also predict 
whether the direct 
neutralino search is likely to produce a signal when the detectors under construction will reach a mass of 10-100kg of Ge 
(in the table the rate per day with 100kg of Ge is indicated\cite{brh}) 
provided that the background can be controlled according to expectations. \par  

To compute the SUSY contribution to g-2=2$\delta a_{\mu}$, I have used the detailed formulae given in \cite{martin}. 
The solution with tan$\beta$=10 is clearly disfavored.
The reason for this is simply that the Higgs mass constraint from LEP2 gives large mass values for
the sneutrino and the chargino which therefore forces large values of tan$\beta$ to satisfy the bound coming from g-2. \par
The choice of tan$\beta\sim$30 is a standard one in grand unified theories with unification of the Yukawa coupling constants of the 3d
generation of fermions. Higher values of 
tan$\beta$ cannot be accomodated with the low values of m$_0$ and m$_{1/2}$ needed for g-2 according to \cite{deboer}.
The conclusion is that with the g-2 result, an e$^+$e$^-$ collider operating 
at 500 GeV can very likely observe sleptons and neutralinos (first + second lightest 
neutralino). \par
 
At 800 GeV there is also access to charginos. Very similar 
numbers can be found in \cite{ellis}. \par
Figure 1 and 2 give a more general picture of the coverage offered by a LC operating at 800 GeV. 
The LSP relic abundance satisfies the cosmological bounds in a small band at the frontier
of the excluded band (co-annihilation with staus with a mass close to the LSP). Combining this cosmological
requirement with the g-2 measurement one can conclude that the preferred region falls well inside the area covered by the
LC for right handed sleptons and of the 2 lightest neutralinos. The coverage for the lightest charginos and left handed sleptons 
is also favoured but with smaller confidence.\par

\begin{table}[t]
\centering
\caption{mSUGRA without the Higgs constraint}
\vskip 0.5 truecm
\begin{tabular}{|c|c|}
\hline
tan$\beta$ & 10 \\
\hline
m$_{1/2}$  & 200     \\
\hline 
m$_0$ & 50  \\
\hline
$\delta a_{\mu}$ x 10$^{10}$  & 54   \\
\hline
M $_{chargino1}$ & 136 \\
\hline
M$_{neutralino1}$ & 71 \\
\hline
M$_{sleptonR}$ &102 \\
\hline
M$_{sneut}$  &137  \\
\hline
M$_{chargino2}$ &284 \\
\hline
M$_{neutralino2}$ &138 \\
\hline
Rate/100kg/day Ge &0.8\\
\hline
\end{tabular}
\end{table} 
These conclusions are based on the mSUGRA approach which is still consistent with the main requirements dictated by the Higgs 
limits/signal from LEP2, g-2 and DM. There are however several motivations to relax the tight unification 
of the mSUGRA scheme and this can be done without contradicting experimental constraints.
Recall for instance that EW 
baryogenesis\cite{wagner} requires
a light stop, of order 100 GeV, which is acceptable if one relaxes the unification of scalar masses at GUT. One could then have a 
light right handed stop and a very heavy left-handed stop (providing the necessary input for the Higgs mass at $\sim$115 GeV). \par
Relaxing the mSUGRA connection between the slepton and the squark masses, 
one could ignore the Higgs limit from LEP2 which 
mainly comes from the stop sector, and allow for lighter chargino/slepton to have a g-2 value consistent with the 
tan$\beta$=10 solution (table 2). Nothing therefore forbids at present the possibility to observe
with e$^+$e$^-$ a 
more complete SUSY spectrum than suggested by the mSUGRA scheme.   \par
Although still speculative, given the present error on g-2 (which however should be reduced by two with the existing data), this 
discussion illustrates the impact of the BNL result. If confirmed, it could herald a brillant future of major
discoveries for a LC operating up to $\sim$1 TeV and for dark matter searches.
                                                    
\vskip 1.5  cm
\centerline {\bf  APPENDIX}
\vskip 1  cm
In this appendix, I will briefly describe the various assumptions made to derive above results. \par
Concerning the derivation of $\mu^2$ from the ESWB relation, 
I have used the approximation valid for large tan$\beta$:  
$$\mu^2\sim 1.5 m^2_{1/2}-0.5M^2_Z-0.04m^2_0$$
As noted in \cite{feng} one can almost neglect the dependence with m$_0$ for the large values of tan$\beta$ considered here. \par
One can in principle also derive m$_A$ using (standard notations):
$$ m^2_A=m^2_1+m^2_2\sim m^2_{H_1}+\mu^2$$
At moderate values of tan$\beta$, one should have :
$$m^2_{H_1}=m^2_0+0.52m^2_{1/2}+\Delta m^2$$
where $\Delta m^2$ is a correction expected to be negative and important at large tan$\beta$. Checking with the results given in 
\cite{ellis} there seems to be good agreement even up to tan$\beta$=30 as 
shown in table 3. This implies that the CP-odd Higgs boson 
will be light enough that one will be able to measure sizeable deviations on the light Higgs branching 
ratios\cite{TDR}. This important conclusion has to be checked more precisely. \par 
\begin{table}[t]
\centering
\caption{Approximate values for m$_A$ }
\vskip 0.5 cm
\begin{tabular}{|c|c|c|c|}
\hline
tan$\beta$ & 10 & 30 & 50 \\
\hline
m$_{1/2}$  & 250   &350   & 400   \\
\hline 
m$_0$ & 100   & 170 & 350  \\
\hline 
m$_A$ approx & 369 &525 & 667\\
\hline
m$_A$ from \cite{ellis}&380& 475& 460 \\
\hline
\end{tabular}
\end{table} 
\vskip 1cm
{\bf Acknowledgements}
 Useful discussions with A. Djouadi and P. Zerwas
are gratefully acknowledged.

\begin{figure}
\epsfysize13cm
\epsfxsize13cm
\epsffile{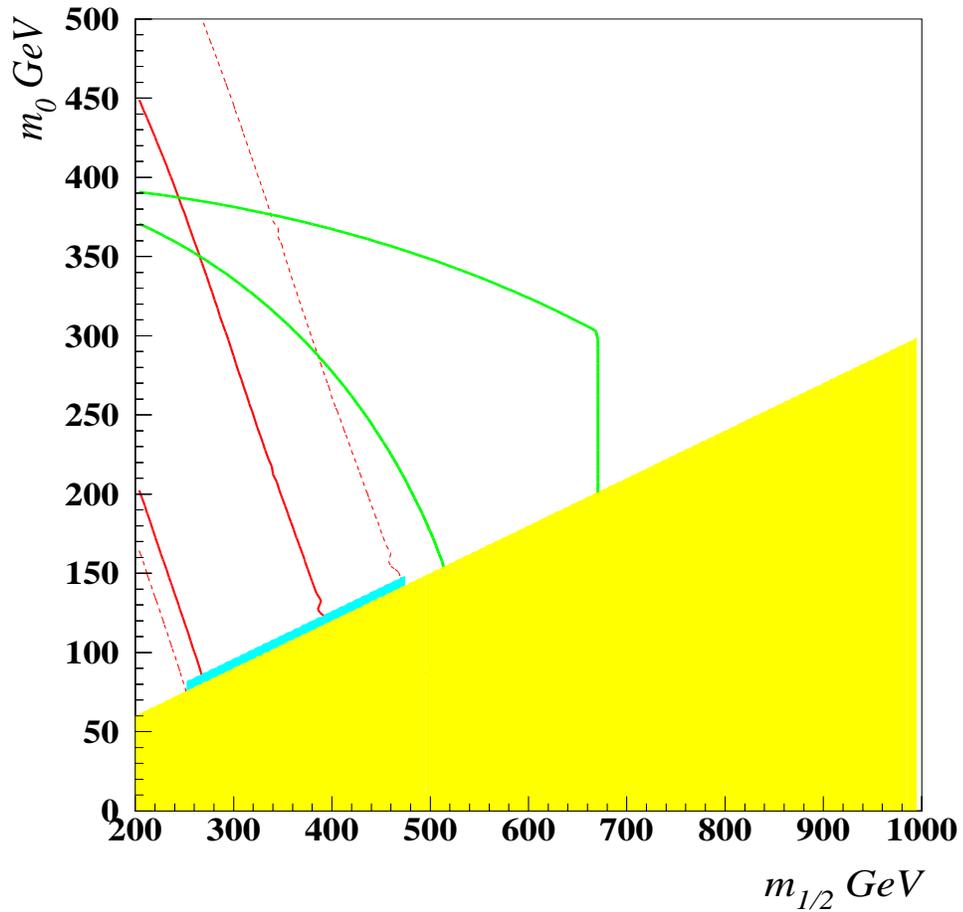}
\caption{g-2 constraints at $\pm$1$\sigma$ (full red curve) and $\pm$1.5$\sigma$ (dashed red curve) for
tan$\beta$=20. The external green contour represents the e$^+$e$^-$  
reach with 800 GeV for right-handed sleptons+neutralino1+neutralino2. The internal green contour also includes the coverage
for left-handed sleptons and the lightest chargino.
The yellow shaded region is excluded by the requirement
that the LSP be neutral. The small blue band is the region with satisfies both the $\pm$1.5$\sigma$ constraint on g-2 and 
gives the correct LSP relic abundance for cosmology (co-annihilation with staus with a mass close to the neutralino LSP)}
\end{figure}
\begin{figure}
\epsfysize12cm
\epsfxsize12cm
\epsffile{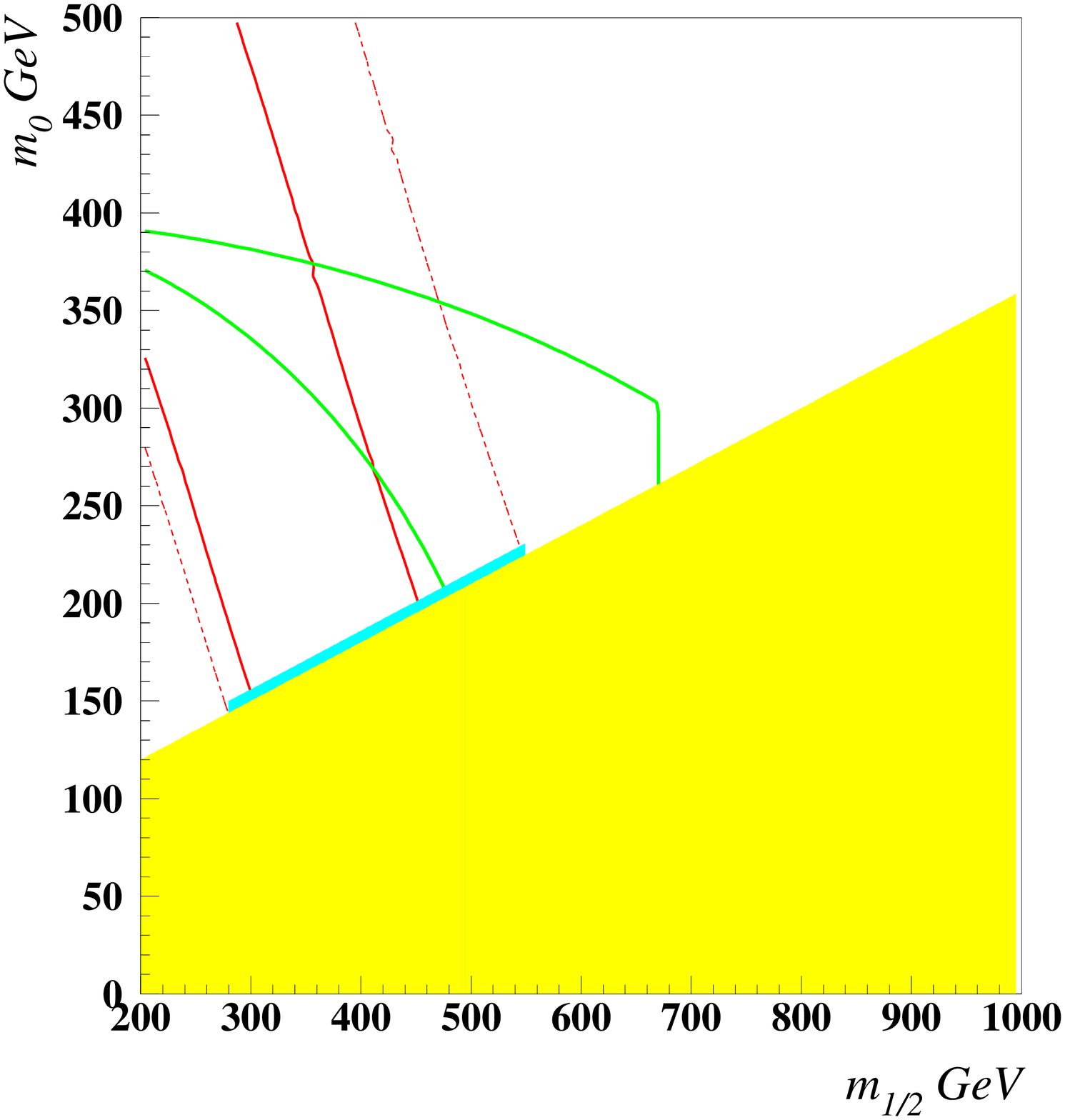}
\caption{g-2 constraints at $\pm$1$\sigma$ and $\pm$1.5$\sigma$ for
tan$\beta$=30. Same conventions as for figure 1.}
\end{figure}


\vskip 1cm
\noindent 
\begin{thebibliography}{99}
\bibitem{fengr}Jonathan L. Feng (Princeton, Inst. Advanced Study), Konstantin T. Matchev (Fermilab), 
Frank Wilczek (Princeton, Inst. Advanced Study). IASSNS-HEP-00-55, FERMILAB-PUB-00-171-T, Aug 2000.
36pp. Published in Phys.Rev.D63:045024,2001 e-Print Archive: astro-ph/0008115
\bibitem{bnl}Muon g-2 Collaboration (H.N. Brown et al.). Feb 2001. 5pp. 
Published in Phys.Rev.Lett.86:2227-2231,2001 
e-Print Archive: hep-ex/0102017 
\bibitem{kane}See for instance
Lisa Everett, Gordon L. Kane, Stefano Rigolin, Lian-Tao Wang (Michigan U.). MCTP-01-02, Feb 2001. 4pp. 
e-Print Archive: hep-ph/0102145 
\bibitem{deboer} W. de Boer, M. Huber (Karlsruhe U.), A.V. Gladyshev, D.I. Kazakov (Dubna, JINR). IEKP-KA-2001-03, Feb 2001. 13pp. 
e-Print Archive: hep-ph/0102163 
\bibitem{brh}
Howard Baer, Michal Brhlik (Florida State U.). FSU-HEP-970605, Jun 1997. 25pp. 
Published in Phys.Rev.D57:567-577,1998 
e-Print Archive: hep-ph/9706509 
\bibitem{martin}
Stephen P. Martin (Northern Illinois U. and Fermilab), James D. Wells (UC, Davis and LBL, Berkeley). FERMILAB-PUB-01-030-T, 
LBNL-47586, Mar 2001. 28pp. 
e-Print Archive: hep-ph/0103067 
\bibitem{ellis}John Ellis (CERN), D.V. Nanopoulos (Texas A-M and HARC, Woodlands and Athens Academy), Keith A. Olive (CERN and 
Minnesota U.).
CERN-TH-2001-054, ACT-02-01, CTP-TAMU-06-01,
UMN-TH-1943-01, TPI-MINN-01-12, Feb 2001. 13pp. 
e-Print Archive: hep-ph/0102331 
\bibitem{wagner}
 C.E.M. Wagner (Argonne National Laboratory, EFI Univ of Chicago)
at the conference 'Higgs and Supersymmetry' Orsay March 19, 2001
 http://ww2.lal.in2p3.fr/actualite/conferences/higgs2001
\bibitem{feng}
Jonathan L. Feng (MIT, LNS). MIT-CTP-3065, Oct 2000. 4pp. 
Talk given at 5th International Linear Collider Workshop (LCWS 2000), Fermilab, Batavia, Illinois, 24-28 Oct 2000. 
e-Print Archive: hep-ph/0012277 
\bibitem{TDR}TESLA TDR, D. Heuer, (ed.), D. Miller, (ed.), F. Richard, (ed.), P. Zerwas, (ed.) (DESY  Hamburg, U.  Orsay, LAL). 
DESY-01-011C, DESY-2001-011C, DESY-TESLA-2001-23C,
DESY-TESLA-FEL-2001-05C, ECFA-2001-209C, Mar 2001. 188pp. 
\end{thebibliography}
\end{document}